\def\mct{\beta_{mc}}
\def\sw{\sigma_W}
\def\fj{\mathcal{J}}
\def\fg{\mathcal{G}}

\def\cone{\mathcal{C}}

\def\cone{\mathcal{C}}

\def\bz{\mathbf{z}}
\def\bZ{\mathbf{Z}}
\def\nb{N}
\def\nc{\mathcal{N}}
\def\ns{N_S}

\def\obs{\bar{\Omega}^2}
\def\te{E}
\def\epp{\mathcal{E}}
\def\intc{\int_{\mathcal{C}}dz}
\def\aa{\mathcal{A}}
\def\bb{\mathcal{B}}
\def\dd{\mathcal{D}}
\def\ww{\mathcal{W}}
\def\zojo{Z[j(\cdot)]}
\def\dos{\Sigma}
\def\bbar{\bar{\beta}}

\documentclass[aps,reprint,pdftex,nofootinbib,superscriptaddress]{revtex4-1}
\usepackage{amsmath,amssymb,amsthm}
\usepackage{graphicx,color}
\usepackage[pdftex,colorlinks]{hyperref}
\usepackage{amsmath}
\usepackage{amssymb}
\usepackage{tipa}
\usepackage{bm}
\usepackage{graphicx}
\usepackage{color}

\begin{document}

\title{Microcanonical work and fluctuation relations for an open system: An exactly solvable model}
\author{Y. Suba\c{s}\i} 
\email{ysubasi@umd.edu}
\affiliation{Joint Quantum Institute and Maryland Center for Fundamental Physics, University of Maryland, College Park, Maryland 20742}
\author{C. Jarzynski}
\email{cjarzyns@umd.edu}
\affiliation{Department of Chemistry and Biochemistry\\
and Institute for Physical Science and Technology,\\
University of Maryland, College Park, Maryland 20742}

\date{\today}

\begin{abstract} 
We calculate the probability distribution of work for
  an exactly solvable model of a system interacting with its
  environment.
  The system of interest is a harmonic oscillator with a
  time dependent control parameter, 
  the environment is modeled by $\nb$ independent harmonic
  oscillators with arbitrary frequencies, and the system-environment
  coupling is bilinear and not necessarily weak.
 The initial conditions of the combined system and
  environment are sampled from a
  microcanonical distribution and the system is driven out of
  equilibrium by changing the control parameter according to a
  prescribed protocol.
  In the limit of infinitely large environment, i.e.
  $\nb\rightarrow \infty$, we recover the nonequilibrium work relation 
  and Crooks's fluctuation relation.
  Moreover, the microcanonical Crooks relation is verified for
  finite environments.
  Finally we show the equivalence of multi-time correlation functions
  of the system in the infinite environment limit for canonical and
  microcanonical ensembles.
\end{abstract}

\maketitle


 \section{Introduction}   
\label{sec:intro}

Recent advances in technology, like real time monitoring and control
of single molecules, enable experiments where small systems can be
studied under nonequilibrium conditions \cite{Bustamante2005}. 
Alongside these advances, there has been considerable progress
in our theoretical understanding of the nonequilibrium
statistical mechanics of small systems.
In this paper we will be concerned in particular with the nonequilbrium
work relation~\cite{Jarzynski1997a,Jarzynski1997b}, 
\begin{align} 
  \langle  e^{- \beta W} \rangle = e^{-\beta \Delta F},
  \label{JE}
\end{align}
and the closely related fluctuation relation, due to Crooks~\cite{Crooks1998,Crooks1999,Crooks2000},
\begin{align} 
  \frac{P(W)}{\tilde{P}(-W)}=e^{\beta(W-\Delta F)} \, .
  \label{CFT}
\end{align}
Both of these relate the statistical fluctuations in the work $W$ performed
on a system during a
nonequilibrium process, to a free energy difference $\Delta F$ between two
equilibrium states of the system. The angular brackets in Eq.~\eqref{JE} denote an
average over an ensemble of realizations of the process, and $\beta$ specifies the inverse temperature
at which the system is prepared prior to each realization.
In Eq.~\eqref{CFT} the numerator and denominator denote the distributions
of work values corresponding to a conjugate pair of ``forward'' and ``reverse'' processes.
Eqs.~\eqref{JE} and \eqref{CFT} have been derived by various means,
using a variety of equations of motion to model the
microscopic dynamics -- see Ref.~\cite{Jarzynski2011} for a review with extensive references -- and have been confirmed
experimentally~\cite{Liphardt2002,Douarche2005,Blickle2006,Douarche2005b,Collin2005,Harris2007}.

We will focus our attention on the formulation of these results within the framework of classical, Hamiltonian dynamics.
The Hamiltonian for the system of interest is assumed to depend on a control parameter
$\lambda$,
whose time dependence over an interval $0\le t\le\tau$ is specified by a schedule, or \emph{protocol}, $\lambda_t$.
The free energy difference $\Delta F$ refers to two different equilibrium
states, corresponding to the initial and final parameter values, $\lambda_0$ and $\lambda_\tau$.

Ref.~\cite{Jarzynski1997a} presents Hamiltonian derivations of Eq.~\eqref{JE} for two different scenarios.
In the first, a system of interest is prepared in equilibrium by being placed in weak contact with a
thermal reservoir, which is then removed.
In this case it is natural to treat the initial conditions of the system of interest as a random sample from the canonical distribution (reflecting the method of preparation),
and also to use Hamiltonian dynamics to model the subsequent evolution of the thermally isolated system as the control parameter is varied ($0\le t\le\tau$).

In the second scenario considered in Ref.~\cite{Jarzynski1997a}, the system remains in weak thermal contact with the reservoir throughout the process.
In this derivation, initial conditions for the combined system \emph{and reservoir} were assumed to be sampled from a canonical distribution,
and then Hamilton's equations were used to model evolution in the full phase space.
In Ref.~\cite{Jarzynski2004} this approach was extended to a system in strong thermal contact with a reservoir, again assuming canonically
sampled initial conditions in the full phase space.

In both derivations described in the previous paragraph, the use of Hamilton's equations to model the dynamics in the full phase space
implies that the combined system of interest and reservoir are being treated as a large, thermally isolated system.
The assumption of a canonical distribution of initial conditions for
this combined system renders the derivation of Eq.~\eqref{JE}
(as well as Eq.~\eqref{CFT}) straightforward.
However, from a conceptual perspective this assumption is somewhat problematic, as the equilibrium state of an isolated system is
typically associated with the {\it microcanonical} ensemble.
It is therefore natural to wonder whether Eqs.~\eqref{JE} and \eqref{CFT} remain valid when initial conditions are sampled microcanonically rather than canonically.
In this paper we will address this question through the exact analysis of a model system, involving a harmonic oscillator (the system of interest) coupled strongly to a bath of $N$ other harmonic oscillators (the thermal reservoir).
This model has previously been studied by Hasegawa \cite{Hasegawa2011}, who considered initial conditions sampled from the canonical ensemble.
More generally, the study of model systems for which exact results can be obtained has illustrated and illuminated a variety of issues related to Eqs.~\eqref{JE} and \eqref{CFT}.~\cite{Mazonka1999,Ritort2002,Ritort2004b,Imparato2005a,Speck2005a,Dhar2005,Lua2005a,Bena2005,Presse2006b,Astumian2007,Crooks2007,Jayannavar2007,Deffner2008,Sung2008,Talkner2008,Deffner2010,Hijar2010,Jimenez-Aquino2010,Morgado2010,Ohzeki2010,Speck2011,Ryabov2013}

It is widely believed that in the thermodynamic limit, the average thermodynamic properties of a
physical system are independent of the choice of the ensemble.
This is the idea of \emph{ensemble equivalence}~\cite{Balian1991}.
However, the situation is quite different when fluctuations are considered~\cite{Lebowitz1967}. 
(As a simple example note that the variance of the total energy is proportional to heat capacity in the canonical ensemble, but vanishes identically in the microcanonical ensemble.)
This suggests that the validity of Eqs.~\eqref{JE} and \eqref{CFT}, for microcanonically sampled initial conditions, does not follow immediately from the equivalence of ensembles, even when the thermal reservoir is assumed to be macroscopic.
This issue is especially relevant since large fluctuations with
very small probabilities play a dominant role in the nonequilibrium
work relation~\cite{Jarzynski2006} whereas standard
ensemble equivalence results do not make any claim about or
depend on such low probability events. Moreover, the work $W$ is not simply a function of the phase space
variables, but rather a functional of the phase space trajectory, and
its fluctuations may be more complex than that of typically considered phase space functions.

For a system interacting with a large environment it has been
suggested in Ref.~\cite{Jarzynski2000}, using heuristic arguments,
that the validity of the nonequilibrium work relation may be insensitive to the particular distribution used and that the canonical ensemble should be
viewed primarily as a computational convenience. 
A more detailed argument supporting this claim has been developed in
Ref.~\cite{Park2004}.
In Ref.~\cite{Cleuren2006} the following microcanonical version of the Crooks fluctuation relation was derived:
\begin{equation}
\frac{P_{\te}(W)}{\tilde{P}_{\te+W}(-W)}=\frac{\dos_f(\te+W)}{\dos_i(\te)},
\label{eq:mcFT}
\end{equation}
where $P_{\te}(W)$ stands for the probability density of doing work $W$
during the forward process and
$\tilde{P}_{\te+W}(-W)$ stands for the probability density of
doing work $-W$ during the time reversed process. The subscript
indicates the energy of the microcanonical distribution from which the
initial conditions are sampled.
The right-hand side is the ratio of two densities of states at different
energies and associated with initial and final Hamiltonians.
(Note that Ref. \cite{Cleuren2006} uses $\Omega$ to denote the
density of states, which we reserve for the system frequency. Thus we
opted to use $\Sigma$ for the density of states instead).
It was then argued in Ref.~\cite{Cleuren2006} that in the appropriate
thermodynamic limit, one recovers Eq.~\eqref{CFT}.
To the best of our knowledge, our paper is the first to explore this issue using a model system for which the work distributions can be computed exactly.

The paper is organized as follows.
The model is introduced in Sec.~\ref{sec:model}.
Exact expressions for the left-hand side of
Eq.~\eqref{JE} are obtained in Sec.~\ref{sec:JE} and for the probability
distribution of work in Sec.~\ref{sec:P(W)}. The validity of nonequilibrium
work relation in the limit of an infinite environment is proven in
Sec.~\ref{sec:Tlimit}. The validity of microcanonical Crooks relation is shown
in Sec.~\ref{sec:mcFT}. Ensemble equivalence in its most general
form is shown in Sec.~\ref{sec:ensequ}. Some technical details of the
derivation are provided in
the Appendix~\ref{app:doublelaplace}.

\begin{figure}[h] 
  \begin{center}
    \includegraphics[width=0.4\textwidth]{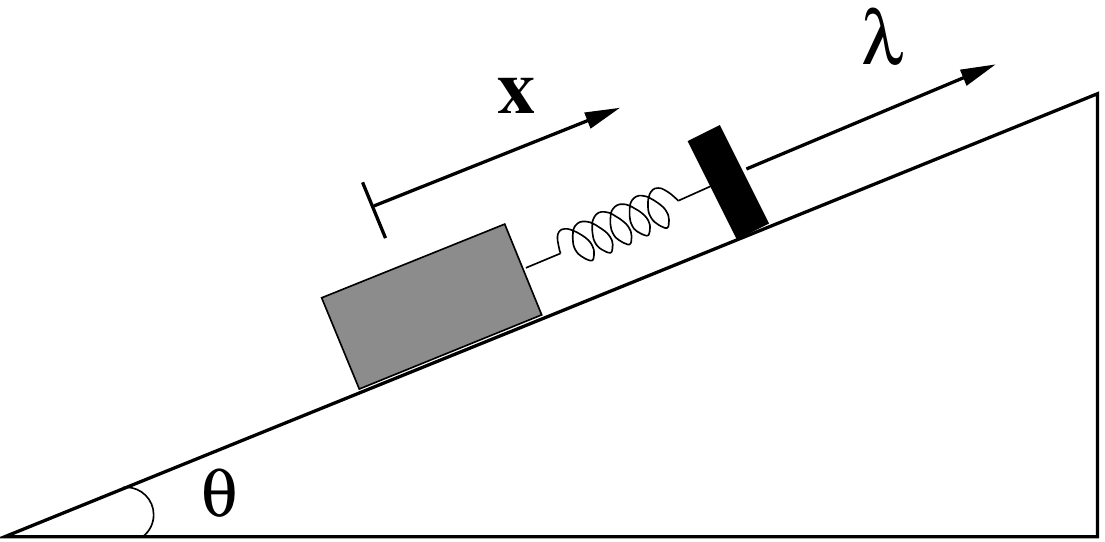}
  \end{center}
  \caption{A mass on a slope is attached to a spring. The support of
  the spring is moved according to a
  time-dependent protocol; $\lambda_t$ denotes the position of the
  support at time $t$. 
  To recover the Hamiltonian \eqref{eq:Heach} one identifies
  $M g \sin\theta \equiv \alpha$.
  Friction
  is modelled via linear coupling to $N$ harmonic oscillators that
  constitute the environment.}
  \label{fig:slope}
\end{figure}

\section{The Model} 
\label{sec:model}

We consider a system of $\nb+\ns\equiv \nc$ classical degrees of
freedom. $\ns$ is the number of degrees of freedom of the system of
interest and $\nb\gg \ns$ is the number of degrees of freedom of the environment. The Hamiltonian governing the dynamics of this closed system is of the form:
\begin{align} 
  \label{eq:Hamiltonian}
  H_{tot}(\bZ,\bz;\lambda)= H_{sys}(\bZ;\lambda) + H_{env}(\bz) +
  H_{int}(\bZ,\bz),
\end{align}
where $\bZ = \{ X_1,P_1,\cdots,X_{\ns},P_{\ns} \}$ and 
$\bz=\{x_{1},p_{1},\cdots,x_{\nb},p_{\nb}\}$, $H_{int}$ is the
interaction Hamiltonian between the system of interest and the
environment, $H_{sys}$ and $H_{env}$ are the system and environment
Hamiltonians respectively. In our model the system
consists of a single harmonic oscillator, i.e. $\ns=1$, and the
environment consists of $\nb$ harmonic oscillators coupled to the
system oscillator bi-linearly:

\begin{align}  
  \nonumber
  H_{sys}(\bZ;\lambda_t) &= \frac{P^2}{2M}+\frac{1}{2}M \Omega^2
(X-\lambda_t)^2 +
\alpha X, \\
\nonumber
H_{int}(\bZ,\bz) &= - \sum_{n=1}^{\nb} c_n x_n X, \\
  \label{eq:Heach}
  H_{env}(\bz) &= \sum_{n=1}^N \left( \frac{p_n^2}{2 m_n}+\frac{1}{2} m_n
\omega_n^2 x_n^2 \right).
\end{align}
Here $\lambda_t$ is a time-dependent parameter determined by
the protocol and $\alpha$ is a constant.
This system Hamiltonian can be realized by the physical system
depicted in Fig.~\ref{fig:slope}. A mass on a slope is attached to a
spring. The support of the spring is moved according to a
  time-dependent protocol; $\lambda_t$ denotes the position of the
  support at time $t$. 
  To recover the Hamiltonian \eqref{eq:Heach} one identifies
  $M g \sin\theta \equiv \alpha$.
  Friction
  is modelled via linear coupling to $N$ harmonic oscillators that
  constitute the environment.
Generalization to more than one system oscillator and allowing for
interactions among environmental oscillators can be achieved
by adopting a matrix notation \cite{Subasi2012b}. However such a general treatment is not
necessary for the purpose of this paper. 

\subsection{The Solution}  

It will prove convenient to define
\begin{align} 
 \label{def:f}
 f(\lambda_t) &\equiv M \Omega^2 \lambda_t- \alpha,\\
  \fj(\lambda_t) &\equiv \frac{1}{2}M \Omega^2 \lambda_t^2.
\end{align}
Then the system Hamiltonian can be written as:
\begin{align} 
  \label{eq:H(f)}
  H_{sys}(\bZ;\lambda_t) &= \frac{P^2}{2M} +\frac{1}{2} M \Omega^2
  X^2-f(\lambda_t)
  X+ \fj(\lambda_t) .
\end{align}

The equation of motion for the system degree of freedom can be
obtained by first solving the dynamics of the environmental degrees of
freedom in terms of the system variables and substituting that
solution into the equation of motion for the system oscillator. The
result is an integro-differential equation for the system oscillator~\cite{Zwanzig1973,Ford1987}
and is referred to as a Langevin equation. 
 \begin{align} 
   \label{eq:langevin}
   \nonumber
 M \ddot{X}(t) &+ 2 M \int_0^t ds\, \gamma(t-s)\dot{X}(s) + M \obs
 X(t)\\
 &=
 f(\lambda_t)- 2M \gamma(t) X(0) + \xi(t), \\
 \label{def:gamma}
 \gamma(t) &\equiv \frac{1}{M} \sum_{n=1}^N \frac{c_n^2}{2 m_n
 \omega_n^2}\cos(\omega_n t), \\
 \label{def:omegabar}
 \obs &\equiv \Omega^2-2\gamma(0),\\
 \label{def:xi}
 \xi(t) &\equiv \sum_{n=1}^N c_n \left(x_n(0) \cos(\omega_n
 t)+\frac{p_n(0)}{m_n \omega_n}\sin(\omega_n t)  \right).
\end{align}
The system-environment coupling is required to satisfy $\Omega^2 \ge 2
\gamma(0)$ for the dynamics to be stable and we will make this
assumption henceforth.

The solution to Eq.~\eqref{eq:langevin} can be written as
\begin{align}  
  \nonumber
  X(t) = &X(0) h(t) + P(0) g(t) + \int_0^t ds\, g(t-s) [
  f(\lambda_s)\\
  &- 2 M
  \gamma(s) X(0) +\xi(s)
  ].
  \label{}
\end{align}
Here $h(t)$ and $g(t)$ are the homogenous solutions of
Eq.~\eqref{eq:langevin} with the right hand side set equal to zero and satisfy
 \begin{eqnarray} 
 h(0)=M\dot{g}(0)=1;\qquad \dot{h}(0)=g(0)=0.
 \end{eqnarray}
The solutions $h(t)$ and $g(t)$ can be calculated using the Laplace transforms:
\begin{equation} 
\label{eq:HGLap}
\hat{h}(s)=\frac{2\hat{\gamma}(s)+s}{s^2+2 s\hat{\gamma}(s)+\obs}, \qquad
\hat{g}(s)=\frac{1/M}{s^2+2s\hat{\gamma}(s)+\obs},
\end{equation}
where the hat indicates Laplace transform. The two linearly
independent homogenous solutions are related by~\cite{Mai2007}:
\begin{align} 
  \label{eq:HGRel0}
s\hat{h}(s)\hspace{-0.4mm}&=\hspace{-0.4mm}1\hspace{-0.4mm}-\hspace{-0.4mm}M\obs\hat{g}(s), \quad
 sM\hat{g}(s)\hspace{-0.4mm}=\hspace{-0.4mm}\hat{h}(s)\hspace{-0.4mm}-\hspace{-0.4mm}2M\hat{\gamma}(s)\hat{g}(s),\\
\label{eq:HGRel}
\dot{h}(t)\hspace{-0.4mm}=\hspace{-0.4mm}&-\hspace{-0.4mm}M\obs g(t), \quad
M\dot{g}(t)\hspace{-0.4mm}=\hspace{-0.4mm}h(t)\hspace{-0.4mm}-\hspace{-0.4mm}2M\hspace{-1.4mm}\int_0^t\hspace{-1.4mm} ds \, \gamma(t-s)g(s).
\end{align}

\section{Nonequilibrium Work Relation} 

We assume a protocol $\lambda_t$ in the time
interval $[0,\tau]$. 
This corresponds to a function $f(\lambda_t)$ via Eq.~\eqref{def:f}.
The work associated with the Hamiltonian \eqref{eq:H(f)} for the duration of the protocol $\Delta t = \tau$ is given by
\begin{align} 
W &= \int_0^\tau dt\, \dot{\lambda} \frac{\partial H_{tot}}{\partial
\lambda} 
=-\int_0^\tau dt\, \dot{f}(\lambda_t)X(t) + \Delta\fj.
\label{eq:work}
\end{align}
The dot over a function indicates time derivative, and $\Delta\fj \equiv
\fj(\lambda_\tau)-\fj(\lambda_0)$. This definition of work is
motivated by the observation $\textrm{\textcrd} W =d\lambda
\frac{H_{tot}}{d \lambda}=\mathrm{displacement}\times
\mathrm{force}$. For a discussion of alternative definitions of work
and various fluctuation theorems they lead to see
Ref.~\cite{Jarzynski2007}.

For the nonequilibrium work relation Eq.~\eqref{JE} the initial state is sampled from the
\emph{canonical} ensemble at inverse temperature $\beta$ using the
Hamiltonian $H_{tot}(\bZ,\bz;\lambda_0)$. 
The free energy
difference is defined via $\Delta F \equiv
F(\lambda_\tau)-F(\lambda_0)$.
In our model the free energies can be calculated explicitly. 
\begin{align} 
  Z_\beta(\lambda) &= e^{- \beta F(\lambda)} = \int d\bZ\, d\bz\, e^{-\beta
  H_{tot}(\bZ,\bz;\lambda)}.
  \label{eq:F1}
\end{align}
Here $Z_\beta(\lambda)$ is the partition function associated with the
Hamiltonian $H_{tot}(\lambda;\bZ,\bz)$.
The integral over the environmental degrees of freedom gives:
\begin{align} 
  \int \hspace{-1.0mm}d \bz\, e^{-\beta(H_{int}(\bZ,\bz)+H_{env}(\bz))}\hspace{-0.4mm} \propto\hspace{-0.4mm}
  e^{\beta X^2 \sum_n\hspace{-0.4mm}\frac{c_n^2}{2 m_n
  \omega_n^2}} = e^{\beta M \gamma(0) X^2}.
\end{align}
Irrelevant constants that will eventually cancel out in the expression
for $\Delta F$ have been omitted in the above expression.
We define the Hamiltonian of mean force as~\cite{Kirkwood1935,Jarzynski2004}:
\begin{align} 
  \nonumber
  H^*(\bZ;\lambda_t)&=H_{sys}(\bZ;\lambda_t)-M \gamma(0) X^2 \\
  &=
  \frac{P^2}{2M}+\frac{1}{2}M \bar{\Omega}^2 X^2-f(\lambda_t) X
  + \fj(\lambda_t),
  \label{eq:H*}
\end{align}
which amounts to shifting the frequency form $\Omega$ to
$\bar{\Omega}$ in the original system Hamiltonian.
Then Eq.~\eqref{eq:F1} becomes (up to some irrelevant constants):
\begin{align} 
 e^{-\beta F(\lambda)} &\propto \int d \bZ\, e^{-\beta
  H^*(\bZ;\lambda)}
  \propto e^{\beta \frac{f(\lambda)^2}{2 M
  \bar{\Omega}^2}-\beta\fj(\lambda)}.
\label{}
\end{align}
The free energy difference is given by
\begin{align} 
   \Delta F &= - \frac{f(\lambda_\tau)^2-f(\lambda_0)^2}{ 2 M
  \bar{\Omega}^2} + \Delta\fj \equiv -\fg+\Delta\fj,
\end{align}
where 
\begin{equation}
  \fg \equiv \frac{f(\lambda_\tau)^2-f(\lambda_0)^2}{2
M \bar{\Omega}^2}.
\end{equation}
Note that an overall shift in $f(\lambda_0)$ simply
changes the equilibrium positions and one is tempted to set
$f(\lambda_0)=0$ in order to simplify the calculation. However, in the
analysis of some fluctuation theorems, where both forward and reverse
processes are considered, this would cause a loss of generality.
Unless $f(\lambda_0) =
f(\lambda_\tau)$, or equivalently $\Delta F =0$, the reverse process is
necessarily described with nonzero $f(\tilde{\lambda}_0)$.

In the next section we will consider the quantity:
\begin{widetext}
\begin{align}  
  \label{eq:JE}
  \langle e^{-\bbar W} \rangle_{\mathrm{mc}} &= \frac{\int d\bZ\,
  d\bz\,
  \delta(H_{tot}(\bZ,\bz;\lambda_0)-\te) \exp\left[\bbar
  \left(\int_0^\tau dt\,
  \dot{f}(t)X(t;\bZ,\bz)-\Delta\fj \right)\right]}{\int d\bZ\, d\bz\,
  \delta(H_{tot}(\bZ,\bz;\lambda_0)-\te)} 
\end{align}
\end{widetext}
which represents the average of $\exp\left( -\bbar W \right)$,
over an ensemble of trajectories with microcanonically sampled initial
conditions in the full phase space. 
We will obtain an exact expression for this average, Eq.~\eqref{eq:jefn} below, valid for any positive value of the parameter $\bbar$.
We will then show that in the thermodynamic limit, $N\rightarrow\infty$, Eq.~\ref{JE} emerges when the value of $\bbar$ is set equal to the inverse temperature $\mct$ associated with the microcanonical energy $E$ (see Eq.~\ref{def:mct}).
That is:
\begin{equation}
\label{eq:recoverNWR}
  \lim_{N\rightarrow \infty} \langle e^{-\mct W} \rangle_{mc} =
  e^{-\mct \Delta F} .
\end{equation}
Although we obtain this result for the case of a single system
oscillator, it is easily generalized to any number $\ns$ of system oscillators, provided the limit $N\rightarrow \infty$ is taken with $\ns$ fixed.
Moreover, heuristic arguments \cite{Park2004} suggest that this result
holds for more general systems with nonlinear interactions.
However nonlinear models are difficult to treat analytically and
careful numerical experiments are necessary to test this hypothesis in
such models. In this work our aim is to  focus on the analytically solvable
harmonic oscillator model, for which exact results can be obtained.

\subsection {Exact Result for finite $N$} 
\label{sec:JE}

The integrals appearing in the denominator and numerator of
Eq.~\eqref{eq:JE} have been computed in Appendix~\ref{app:main}. 
The trick is to use an integral representation of the delta
function in order to transform the integrals over the phase space variables
into Gaussian integrals. Once the phase space integrals are performed,
it is seen that the integration left over from the representation of the delta
function can also be performed exactly. Below we cite the results and
refer the reader to Appendix~\ref{app:main} for the technical details.

Combining Eq.~\eqref{eq:denum} for the denominator and
Eq.~\eqref{eq:num} for the numerator of Eq.~\eqref{eq:JE} we obtain for Eq.~\eqref{eq:JE}: 
  \begin{align}  
    \langle e^{-\bbar W}\rangle_{\mathrm{mc}} =e^{-\bbar \Delta\fj}
    e^{\bbar \fg -
    \bbar^{-1} \dd} \frac{
    N!}{(\aa \dd)^{N/2}}I_N(\sqrt{4\aa \dd}).
    \label{eq:jefn}
  \end{align}
where
  \begin{align}
  \label{def:aa2}
  \aa &\equiv \te + \frac{f(\lambda_0)^2}{2
M \obs}-\fj(\lambda_0),\\
  \label{def:dd2}
  \dd &\equiv
\frac{\bbar^2}{ M \obs} I_f, \\
 I_f &=  \int_0^\tau dt\, \int_0^t ds\,  \dot{f}(\lambda_t)
  h(t-s)\dot{f}(\lambda_s). 
  \label{eq:I_f2}
\end{align}
Eq.~\eqref{eq:jefn} is the exact expression for a system of one harmonic oscillator
dragged up a slope in the presence of gravity and
coupled to an environment modelled by N harmonic oscillators in a microcanonical ensemble at
energy $\te$.

The effect of the environment is implicit in Eq.~\eqref{eq:jefn}. The
microcanonical temperature and $\aa$ both depend on the total
energy $\te$. 
 Also $I_f$ depends on 
$h(t)$, which is the homogenous solution to
the Langevin equation. Finally $\dd$ and $\aa$ contain factors of
$\bar{\Omega}$ which is the renormalized frequency. 

\subsection{The thermodynamic limit, $N\rightarrow \infty$}
\label{sec:Tlimit}

In this limit we define energy per particle
\begin{align} 
  \epp &\equiv \frac{\te}{N+1}= \frac{\te}{N}+O(N^{-1}), \\
  \aa &= N\left( \epp + \frac{1}{N}\left(\frac{f(\lambda_0)^2}{2 M \obs }-
  \fj(\lambda_0) \right) \right)= N \epp +
  O(1).
  \label{eq:aadef}
\end{align}
Eq.~\eqref{eq:jefn} becomes:
   \begin{align} 
    \langle e^{-\bbar W}\rangle_{\mathrm{mc}} =e^{-\bbar \Delta\fj}
    e^{\bbar \fg-
    \bbar^{-1} \dd}\frac{
    N!}{(\epp \dd N)^{N/2}} I_N(\sqrt{4 N \epp \dd}).
    \label{eq:jefn2}
  \end{align}
The asymptotic behaviour of the Bessel function $I_N(x)$ is usually
given for cases where $x$ goes to zero or infinity while $N$ is
fixed. In Eq.~\eqref{eq:jefn2} $x\sim
\sqrt{N}$ as $\nb \rightarrow \infty$. 
Luckily there is a formula for the limit we are looking
for:\footnote{To see this, first note that the Bessel function can be
written in terms of the generalized hypergeometric functions as
$I_N(y)={}_0F_1(N+1;y^2/4)\, (y/2)^N /N!$. Looking at the series expansion
of the hypergeometric function it is easy to see that ${}_0F_1(N,Ny^2/4)
\rightarrow {}_0F_0(y^2/4)$ as $N\rightarrow \infty$. Finally one
notes
that ${}_0F_0(y)=e^y$ to arrive at the desired formula.}
\begin{align}   
  \underset{N \rightarrow \infty}{\mathrm{lim}} I_N(\sqrt{N x})=
  \frac{1}{N!}\left( \frac{N x}{4} \right)^{N/2}e^{x
  /4}.\,  
  \label{eq:diflim}
\end{align}
Using this formula with $x=4\dd \epp$ in Eq.~\eqref{eq:jefn2} we obtain
\begin{align} 
  \underset{N \rightarrow \infty}{\mathrm{lim}}\hspace{-1.2mm} \langle e^{-\bbar
  W}\rangle_{\hspace{-0.3mm}\mathrm{mc}}\hspace{-0.3mm}=\hspace{-0.3mm} e^{-\bbar(\Delta\fj- \fg)-
 \bbar^{\hspace{-0.2mm}-1}\hspace{-0.3mm}\dd+\epp \dd}\hspace{-0.6mm} =\hspace{-0.5mm} e^{-\bbar \Delta F+(\epp-\bbar^{\hspace{-0.2mm}-1}\hspace{-0.3mm} )\dd},
  \label{eq:final}
\end{align}
which, like Eq.~\eqref{eq:jefn}, is valid for arbitrary $\bbar>0$.

Since the quantity $\dd$ depends on the protocol used to vary the parameter $\lambda$ (see Eqs.~\eqref{def:dd2}, \eqref{eq:I_f2}), the right side of Eq.~\eqref{eq:jefn} generally cannot be expressed in terms of a difference between two state functions.
However, consider the particular choice  
\begin{align}
  \label{def:mct}
  \bbar = \mct \equiv \epp^{-1} ,
\end{align}
corresponding to the inverse temperature given by the equipartition theorem for a collection of one-dimensional harmonic oscillators.
For this choice the protocol dependent term vanishes, and -- as advertised (Eq.~\eqref{eq:recoverNWR}) -- we recover the nonequilibrium work relation.

For more general models there is still going to be a well-defined relationship
between energy per particle and temperature, but it will no longer be linear as in Eq. \eqref{def:mct}. In such models we expect
Eq. \eqref{eq:final} will be protocol independent only for the particular choice
of $\bbar=\mct$ which satisfies the corresponding relationship between energy per particle and temperature.

This concludes the derivation of work fluctuation relation 
for a system plus environment Brownian motion model in the microcanonical
ensemble.

\section{Crooks's Fluctuation Relation}
\subsection{Probability Distribution of Work }
\label{sec:P(W)}

The moment generating function of work is defined as:
\begin{align}
  G_W(s) = \langle e^{-i s W}\rangle_{mc}.
  \label{}
\end{align}
It can be obtained from Eq.~\eqref{eq:jefn} by analytic continuation
via $\bbar\rightarrow i s$. The probability distribution of work is the Fourier
transform of the moment generating function
\begin{align}  
  P_{\te}(W) = \frac{1}{2\pi} \int_\cone ds\, e^{is W} G_W(s),
  \label{eq:connect}
\end{align}
where $P_{\te}(W)$ has been defined earlier in the Introduction.

Assuming $I_f>0$ and after some manipulations we are lead to the following formula:
\begin{align}
  P_{\te}(W) = \frac{\nb !\, 2^N}{2\pi \sqrt{\nb 2 \sw^2}} \int_\cone
  ds\, \frac{ e^{is
  \frac{(\ww-\langle W \rangle)}{\sqrt{N 2 \sw^2}}} J_N(s)}{s^N},
  \label{}
\end{align}
where $\langle W \rangle \equiv \Delta F +\frac{I_f}{M \obs}$ is the
expectation value of work and $\sw^2 \equiv 2 \aa I_f/ N M
\obs$ is related to the variance of work in the canonical ensemble,
as we will see later. 
The integral can be done analytically to give:
\begin{align} 
  \nonumber
  P_{\te}(W) = &\frac{\nb
  !}{\Gamma(\nb+1/2)\nb^{1/2}}\frac{1}{\sqrt{2 \pi \sw^2}}\\
  \nonumber
  &\times \left( 1-
  \frac{(W-\langle W\rangle)^2/2 \sw^2}{\nb}
  \right)^{N-1/2} \\
  &\times \Theta\left( \sqrt{2 N} \sw  -|W-\langle W \rangle|
  \right).
  \label{eq:P(W)}
\end{align}
This is the exact expression for the probability distribution of work
done on a single harmonic oscillator coupled to an environment of
$\nb$ harmonic oscillators.

The step function in Eq.~\eqref{eq:P(W)} shows that the maximum
deviation from the average value of work scales as the square root of
$\nb$. The fact that the work is bounded is a consequence of the fact
that microcanonical ensemble describes a distribution with finite
support over the phase space. By applying the method of Lagrange
multipliers on the expression of work \eqref{eq:work}, with the constraint of fixed energy , the extreme values of
work can be verified independently. This analysis also yields
analytical expressions for the phase space trajectory of each particle for the realizations corresponding
to extreme values of work.

The special case of $I_f=0$ is very easy to handle. Using
Eq.~\eqref{eq:I_f=0JE} with $\bbar\rightarrow is$ in Eq.~\eqref{eq:connect} we see that the resulting integral is the representation of the delta function. Hence
$P_{\te}(W)=\delta(W-\Delta F)$ for $I_f=0$.

Next we take the limit of infinite environment. The first factor of
Eq.~\eqref{eq:P(W)} can easily be seen to converge to one as $\nb
\rightarrow \infty$. For the third factor we use the formula:
\begin{align} 
  e^x=\underset{N\rightarrow \infty}{\mathrm{lim}}\left[ 1 +
  \frac{x}{N}
  \right]^N.
  \label{eq:limofexp}
\end{align}
Thus for the infinite environment limit we recover the Gaussian form:
\begin{align}  
  \label{eq:PofW}
  \underset{\nb\rightarrow\infty}{\mathrm{lim}} P_{\te}(W) &=
  \frac{e^{-\frac{(W-\langle W\rangle)^2}{2 \sw^2}}}{\sqrt{2\pi \sw^2}}, \\
   \lim_{N\rightarrow \infty} \sw^2 &= 2 \epp \frac{I_f}{ M
  \obs} = \frac{2}{\mct} \left( \langle W \rangle -\Delta F \right).
 \label{gauss}
\end{align}
Eqs. \eqref{eq:PofW} and \eqref{gauss} ensure that the nonequilibrium work and Crooks's
fluctuation relations are satisfied whenever $\beta$ is identified with 
$\mct$ in Eqs.~(\ref{JE},\ref{CFT})~\cite{Subasi2012a}.
The probability distribution \eqref{eq:PofW} is identical to the probability distribution of work for the case
where the initial conditions of the system plus environment are sampled from a canonical
ensemble, with the temperature of the canonical ensemble related to the total energy of the
microcanonical ensemble according to $\beta =\mct$. This can be easily checked, since all the
integrations are Gaussian for the system plus environment canonical initial
conditions (as opposed to the presence of the delta function in the
microcanonical initial conditions).

\subsection{Microcanonical Crooks Relation }
\label{sec:mcFT}

Below we will show the validity of Eq.~\eqref{eq:mcFT} for our specific
model.
First we note that the initial density of states $\dos_{i}$ is given by the
denominator of Eq.~\eqref{eq:JE}, and a similar expression applies to the final density of states $\dos_{f}$ (only with $\lambda_0$ replaced by $\lambda_\tau$).
From Eq.~\eqref{eq:denum} we have:
\begin{equation} 
  \frac{\dos_f(\te+W)}{\dos_i(\te)} = \left(
  \frac{\tilde{\aa}}{\aa} \right)^N \label{eq:rat1},
\end{equation}
where
\begin{equation}
   \tilde{\aa} \equiv (\te+W)+ \frac{f(\lambda_\tau)^2}{2 M \obs}-\frac{1}{2}M
  \Omega^2 \lambda_\tau^2.
\end{equation}
The expressions for $\dos_f(\te+W)$ and $\tilde{\aa}$ for the reverse process have been 
obtained from Eqs.~\eqref{eq:denum} and \eqref{eq:aadef} by letting
$\lambda_0\rightarrow \lambda_\tau$ and $\te\rightarrow\te+W$. The
probability distribution of work in the forward and reverse processes
are given by:
\begin{align}
  \nonumber
  P_{E}(W)\propto&\frac{1}{\sigma_{W}^{2N}}\left(\frac{2N\sigma_{W}^{2}-\left(W-\langle
W\rangle\right)^{2}}{2N}\right)^{N-1/2}\\
&\times \Theta\left(\sqrt{2N}\sigma_{W}-|W-\langle
W\rangle|\right),\\
\nonumber
\tilde{P}_{E+W}(-W)\propto&\frac{1}{\tilde{\sigma}_{W}^{2N}}\left(\frac{2N\tilde{\sigma}_{W}^{2}-\left(-W-\langle
\tilde{W}\rangle\right)^{2}}{2N}\right)^{N-1/2}\\
&\times \Theta\left(\sqrt{2N}\tilde{\sigma}_{W}-|-W-\langle
\tilde{W}\rangle|\right), \label{eq:Ps}
\end{align}
where
$\tilde{\sigma}_{W}^{2}=2I_{f}\tilde{\mathcal{A}}/NM\bar{\Omega}^{2}$
and the following quantities for the time reversed process have been
defined in analogy with the forward process:
\begin{align}  
   \tilde{\langle W\rangle} &\equiv -\Delta F + \frac{I_f}{M \obs} = \langle
  W\rangle -2 \Delta F = \frac{I_f}{M \obs}-\Delta F,\\
  \Delta\tilde{F} &\equiv -\Delta F= \frac{f(\lambda_\tau)^2-f(\lambda_0)^2}{2 M
  \obs}-\frac{M \Omega^2 \left( \lambda_\tau^2-\lambda_0^2
  \right)}{2},\\
  \tilde{\sigma}_W^2 &\hspace{-0.4mm}=\hspace{-0.4mm} \frac{2 I_f}{M \obs N}\left(\te\hspace{-0.4mm}+\hspace{-0.4mm}W\hspace{-0.4mm}+\hspace{-0.4mm} \frac{f(\lambda_\tau)^2}{2 M \obs}\hspace{-0.4mm}-\hspace{-0.4mm}\frac{1}{2}M
  \Omega^2 \lambda_\tau^2  \right) \hspace{-0.4mm}=\hspace{-0.4mm} \frac{\tilde{\aa}}{\aa}
  \sigma_W^2.
  \label{}
\end{align}
Here we have used the fact that $I_f$ is the same
for the forward and reverse process by the virtue of the symmetry of
its defining double integral. Based on these formulas we can write the
left-hand side of Eq.~\eqref{eq:mcFT} purely in terms of
$\lambda$ and $I_f$, whereas the right-hand side is simply given by
Eq.~\eqref{eq:rat1}. Ignoring the step functions for the moment
Eq.~\eqref{eq:mcFT} can be written as
\begin{align} 
  \nonumber
  \frac{P_E(W)}{\tilde{P}_{E+W}(-W)} &= \left(
  \frac{\tilde{\sigma}_W^2}{\sigma_W^2} \right)^N \left(
  \frac{2N \sigma_W^2-\left( W-\langle W \rangle \right)^2}{2 N
  \tilde{\sigma}_W^2-(-W-\langle\tilde{W}\rangle )^2}
  \right)\\
  \nonumber
  &= \left(
  \frac{\tilde{\aa}}{\aa} \right)^N \left(
  \frac{2N \sigma_W^2-\left( W-\langle W \rangle \right)^2}{2 N
  \tilde{\sigma}_W^2-(-W-\langle\tilde{W}\rangle )^2}
  \right) \\
  &= \left( \frac{\tilde{\aa}}{\aa} \right)^N,
  \label{}
\end{align}
This implies, again disregarding the step function for the moment,

\begin{align} 
  &2N \sigma_W^2-\left( W-\langle W \rangle \right)^2 = 2 N
  \tilde{\sigma}_W^2-(-W-\langle\tilde{W}\rangle )^2,\\
  &2 W \left( \langle W \rangle\hspace{-0.4mm}+\hspace{-0.4mm}\langle \tilde{W} \rangle
  \right)\hspace{-0.4mm}+\hspace{-0.4mm}\left( \langle \tilde{W}\rangle^2\hspace{-0.4mm}-\hspace{-0.4mm}\langle W\rangle^2 \right)
  =  2N \left( \tilde{\sigma}_W^2 \hspace{-0.4mm}-\hspace{-0.4mm}\sigma_W^2 \right),
  \label{eq:coincidence}
\end{align}
This equality can be verified by calculating the following relations.
\begin{align} 
  \langle W \rangle +\langle \tilde{W} \rangle &=
  \frac{2 I_f}{M \obs}, \\
  \langle \tilde{W} \rangle^2 - \langle W \rangle^2 &=
  -\frac{4 I_f}{ M \obs} \Delta F,\\
  \tilde{\sigma}_W^2 -\sigma_W^2 &= \frac{2 I_f}{M \obs N}(W- \Delta
  F).
  \label{}
\end{align}

Now we return to the question of whether the step functions appearing
in $P_{\te}(W)$ and $P_{\te+W}(-W)$ are identical, so that they cancel
when forming the ratio Eq.~\eqref{eq:mcFT}.
To this end consider the conditions for the probabilities $P_E(W)$ and
$\tilde{P}_{E+W}(-W)$ to vanish:
\begin{align} 
  2N \sigma_W^2 &= \left( W-\langle W\rangle \right)^2, \\
  2N \tilde{\sigma}_W^2 &= \left( W+\langle \tilde{W}\rangle
  \right)^2.
  \label{}
\end{align}
To see that both conditions are identical observe that the difference
of both equations gives Eq.~\eqref{eq:coincidence} which has been shown to
hold. Thus we have demonstrated the validity of the microcanonical
Crooks relation in our particular model.

\section{Ensemble Equivalence} 
\label{sec:Equivalence}

In most textbooks the term \emph{ensemble equivalence} is used to
describe the following property of extensive systems: macroscopic
physical quantities assume the same value in any equilibrium ensemble,
i.e. microcanonical, canonical or grand canonical. In this section we
will deviate from this definition in three ways. The system plus
environment model considered
in this paper is not extensive. Second, the thermodynamic limit is taken with
the system size fixed (in the particular case treated here the system
consists of a single oscillator). Thus the quantities we consider do
not have to be
macroscopic. Third, we will consider multi-time averages taken over
nonequilibrium processes.

\subsection{Initial Phase Space Distribution} 
\label{sec:iniphaspa}

In this section we show that as $\nb\rightarrow \infty$ the
phase space probability density of the system oscillator approaches that of a
canonical distribution if the probability distribution for the system plus
environment closed system is given by the microcanonical distribution. 

The derivation is similar to the previous sections. 
\begin{align} 
  \mathsf{f}_S(\bZ) &= \frac{\int d\bz\,
  \delta(H_{tot}(\lambda;\bZ,\bz)-\te)}{\int d\bZ\, d\bz\,
  \delta(H_{tot}(\lambda;\bZ,\bz))}.
  \label{}
\end{align}
For the numerator we again substitute the integral representation of
the delta function to obtain:
\begin{align} 
  \nonumber
  \int_\cone dz\, e^{-i z \te} \int d\bz\, e^{i
  (H_{sys}(\bZ)+H_{int}(\bZ,\bz)+H_{env}(\bz))} \\
  = \int_\cone dz\, \frac{e^{-i z
  (\te -H^*(\lambda;\bZ))}}{z^N}.
  \label{eq:P_S}
\end{align}
Here we used Eq.~\eqref{eq:intout}. This integral can be
obtained using the Cauchy theorem. The integrand has a pole of order
$N$ at the origin and the integration contour $\cone$ is passing below
this pole in the complex plane. For $\te > H^*(\lambda;\bZ)$ the
contour can be closed from above to enclose the pole, and there is a
nonzero outcome. For $\te < H^*(\lambda;\bZ)$ the contour is closed
from below where the function is analytic. Hence the outcome of the
integral is zero. 
The final expression for the normalized probability density of system degrees of
freedom is given by:
\begin{align}   
  \mathsf{f}_S(\bZ) = \frac{N \bar{\Omega}}{2 \pi}
  \frac{(\te-H^*(0;\bZ))^{N-1}}{\aa^N}\Theta(\te-H^*(0;\bZ)).
  \label{eq:notyetBoltzmann}
\end{align}
where $\Theta$ denotes the Heaviside step function. The existence of the step
function is a manifestation of the fact that the energy of the system
oscillator cannot exceed that of the system plus environment.

Next consider the $N\rightarrow \infty$ limit. 
\begin{align} 
  \nonumber
  \underset{N \rightarrow \infty}{\mathrm{lim}} \mathsf{f}_S(\bZ) &=
  \frac{\bar{\Omega}}{2\pi \epp} \underset{N \rightarrow \infty}{\mathrm{lim}}\left( 1-\frac{H^*(\bZ)/\epp}{N+1}
  \right)^{N-1}\\
  &=\frac{\bar{\Omega}}{2\pi \epp}e^{-\epp^{-1} H^*(\bZ)},
  \label{eq:limofP_S}
\end{align}
where we used \eqref{eq:limofexp} in the last equality.
The limit in Eq.~\eqref{eq:limofP_S} needs to be interpreted as follows: For
any finite $N$ the probability density \eqref{eq:notyetBoltzmann} agrees with the
canonical distribution \eqref{eq:limofP_S} for small energies. However at
large enough energies relative differences become significant. These
differences would also show up at high order moments of position and
momenta. The limit in Eq.~\eqref{eq:limofP_S} means that given an
energy interval or
equivalently a maximum order for the moments of interest, one can choose a
large enough $N$ such that the microcanonical result will agree with
the asymptotic result to the desired degree.

Eq.~\eqref{eq:limofP_S} describes a Boltzmann state with the Hamiltonian of mean force
replacing the system Hamiltonian. 
Note that the same probability distribution is obtained, albeit for any
$\nb$, if the system plus environment is sampled from a canonical
distribution. In fact this is how the Hamiltonian of mean force is usually
motivated. Eq.~\eqref{eq:limofP_S} states that for a large environment
the phase space density of the system degrees of freedom is the same if the system plus environment is sampled from
a canonical or microcanonical distribution.

\subsection{Multi-time Correlations}
\label{sec:ensequ}

The most general multi-time correlation function during the
nonequilibrium process can be obtained from the generating functional
\begin{align} 
  Z_{ens}[j(\cdot)] &= \langle e^{\int_0^\tau dt\, j(t)
  X(t)} \rangle_{ens},
  \label{}
\end{align}
where $X(t)$ is the solution to the equations of motion with some
initial conditions and the
averaging is done over the desired ensemble. Here we will compare the
generating functionals for the canonical and microcanonical ensembles.
Any multi-time correlation can be obtained from the generating
functional by applying differential operators to it, for example:
\begin{align} 
  \frac{\delta}{\delta j(t_1)} \zojo \Big|_{j=0} &= \langle X(t_1)
  \rangle, \\
  M \frac{\partial}{\partial t_1}\frac{\delta}{\delta j(t_1)} \zojo
  \Big|_{j=0} &=
  \langle P(t_1) \rangle, \\
  \nonumber
  M\hspace{-0.4mm} \frac{\partial}{\partial t_1\hspace{-0.4mm}}\frac{\delta}{\delta j(t_1\hspace{-0.4mm})}\cdots M\hspace{-0.4mm}
  \frac{\partial}{\partial t_k}\frac{\delta}{\delta j(t_{k}\hspace{-0.4mm})}&
  \frac{\delta}{\delta j(t_{k+1}\hspace{-0.4mm})}\cdots\frac{\delta}{\delta
  j(t_{l}\hspace{-0.4mm})}\zojo \Big|_{j=0} \\
  = \langle P(t_1)\cdots P(t_k) &X(t_{k+1}) \cdots
  X(t_l))\rangle.
  \label{}
\end{align}
Note that even the average appearing in nonequilibrium work relation
Eq.~\eqref{eq:JE} can be obtained from this generating functional via
\begin{align} 
  \langle e^{-\beta W} \rangle = e^{-\beta \Delta \fj} \langle
  e^{\beta \int_0^\tau dt\, \dot{f}(t) X(t)} \rangle = e^{-\beta \Delta
  \fj} Z[\beta \dot{f}(\cdot)].
  \label{}
\end{align}
The results presented in this section thus include that of
Sec.~\ref{sec:JE} as a sub-case.

The calculation of the generating functional in both canonical and
microcanonical ensembles
is straightforward but tedious. For the canonical ensemble the
calculation involves only Gaussian integrals
and the use of properties of the solutions of the Langevin equation. The
derivation for the microcanonical ensemble mimic closely the treatment
presented in Appendix~\ref{app:main}. 
Here we only provide the final results.
\begin{align}
 \nonumber
  &Z_{can}[j(\cdot)] = e^{\int_0^\tau\hspace{-0.5mm} dt\,\hspace{-0.5mm} j(t) \left(
  \frac{f(\lambda_0)}{M \obs} h(t) + \int_0^t\hspace{-0.5mm}
  ds\,\hspace{-0.5mm} g(t-s) f(\lambda_s) \right)} \\
  &\hspace{3cm} \times e^{\int_0^\tau\hspace{-0.5mm} dt\,\hspace{-0.5mm} \int_0^t\hspace{-0.5mm} dt'\,\hspace{-0.5mm} j(t) \left(
  \frac{h(|t-t'|)}{\beta M \obs} \right)j(t')} \\
  &= \exp\left(\hspace{-0.5mm} \int_0^\tau\hspace{-1.7mm} dt\,\hspace{-0.5mm} j(t) \langle X(t) \rangle + \int_0^\tau\hspace{-1.7mm}
  dt\,\hspace{-0.9mm} \int_0^t\hspace{-1.7mm} dt'\,\hspace{-0.5mm} j(t) \sigma_{xx}(t,t') j(t')\hspace{-0.5mm} \right), \\
  &Z_{mc}[j(\cdot)] = \exp\left( \int_0^\tau\hspace{-0.5mm} dt\,\hspace{-0.5mm} j(t) \langle X(t)
  \rangle \right) \nonumber \\
  &\hspace{2cm} \times 
  \frac{N!}{\left(\aa \bar{\dd}[j(\cdot)]\right)^{N/2}} I_N\left(
  \sqrt{4 \aa \bar{\dd}[j(\cdot)]}
  \right).
  \label{eq:Z[j]}
\end{align}
where $\langle X(t)\rangle$ stands for the average
position at time $t$ and $\sigma_{xx}(t,t')\equiv \langle X(t) X(t')\rangle - \langle
X(t) \rangle \langle X(t') \rangle$ stands for the two time
fluctuations of the position. We also defined
$\bar{\dd}[j(\cdot)]\equiv \int_0^\tau dt\, \int_0^t dt'\, j(t)
\frac{h(t-t')}{M \obs} j(t')$ analogous to $\dd$ whereby $j(t)$
replaces $\beta \dot{f}(\lambda_t)$.

The equivalence of $Z_{can}$ and $Z_{mc}$ in the $N\rightarrow
\infty$ limit for fixed $j(\cdot)$ follows directly form the
asymptotic formula of the Bessel function given by Eq.~\eqref{eq:diflim}.
\begin{align}
  \lim_{N\rightarrow \infty} Z_{mc}[j(\cdot)] = Z_{can}[j(\cdot)]
  \label{lim}
\end{align}
Similar to the discussion at the end of the previous section the
meaning of this limit calls for some elaboration. As mentioned before
the generating functional can be used to obtain correlation functions.
For large but fixed $N$ and given force protocol and temperature, the low order correlation functions for
microcanonical and canonical ensembles will be very close. However one
can always go to high enough orders where relative differences will
become significant. The limit in Eq.~\eqref{lim} means that given a
certain order we can always choose a large enough $N$ such that the
microcanonical correlation functions up to that order agree with the
corresponding canonical correlation functions to the desired degree.

\section{Discussion}
\label{sec:discussion}

In this paper we treated the exactly solvable model of a harmonic
oscillator driven out of equilibrium by an external force and bilinearly coupled to an
environment of $\nb$ harmonic oscillators. An exact expression for
the probability distribution of work, i.e. Eq.~\eqref{eq:P(W)}, is obtained for any value of
$\nb$, assuming that the combined system
and environment is initially sampled from the microcanonical ensemble. 
Using this expression the microcanonical Crook's relation
\eqref{eq:mcFT} is verified.
In the limit of an infinite environment, nonequilibrium work relation 
\eqref{JE}
and Crooks's fluctuation relation \eqref{CFT} are shown to hold. 
Finally in Sec.~\ref{sec:ensequ} the equivalence of
all multi-time correlations of the system oscillator in the
canonical and microcanonical ensembles in the infinite environment
limit is obtained. 

Our results support the hypothesis that for macroscopically large
environments the sampling of the initial conditions from a canonical
or microcanonical distribution is equivalent as far as system
observables are concerned. 

In the model used in this paper the system oscillator is singled out
not just by the virtue of the time-dependent force being only applied to it
but also by the fact that all the environmental modes are coupled to
it but not to each other. 
This may seem like a limitation of the model. However, the most general system of coupled harmonic oscillators, i.e. allowing for the
environmental oscillators to couple among themselves, can be
represented by the model used in this paper by first decomposing the
environment into its eigenmodes, which in turn leads to a trivial change in the
environment frequencies $\omega_n$ and coupling constants $c
_n$ \cite{Ford1988}. Since
we allow for arbitrary $\omega_n$ and $c_n$ in our derivation, our model is able to
represent any set of coupled harmonic oscillators.

\appendix

\section{Derivation of the Main Result Eq.\eqref{eq:JE}}
\label{app:main}

In this appendix we will compute the integrals appearing in
Eq.~\eqref{eq:JE}. But first we review the integral representation of
the delta function to be used in the derivation.

\subsubsection{The Delta Function}

The delta functions make the integrals in \eqref{eq:JE}
difficult to evaluate. To get around this difficulty we invoke the
following integral representation of the delta function:
\begin{align}
  \label{deltam}
\delta(H_{tot}(\bZ,\bz;\lambda_0)-\te) = \frac{1}{2\pi}
\int_{-\infty}^{\infty} ds\, e^{-i s (H_{tot}(\bZ,\bz;\lambda_0)-\te)}.
\end{align}
The logic behind this is to convert the phase space integral into a simple Gaussian integral. After we perform that integral we will be able to do the $s$ integration as well.

Observe that the integral formula for the delta function can be
modified by allowing the integration variable $s$ to have a constant
imaginary part. We rename it $z$ to emphasize the complex nature:
\begin{align}  
  \nonumber
\frac{1}{2\pi} &\int_{-\infty-i \epsilon}^{\infty-i\epsilon} dz\, e^{-i z
(H_{tot}(\bZ,\bz;\lambda_0)-\te)} \\
&= \frac{1}{2\pi}
\int_{-\infty}^{\infty} ds\,
e^{-i (s-i\epsilon) (H_{tot}(\bZ,\bz;\lambda_0)-\te)}\\
&=e^{-\epsilon (H_{tot}(\bZ,\bz;\lambda_0)-\te)}
\frac{1}{2\pi}\int_{-\infty}^\infty ds\, e^{-i s \left(
H_{tot}(\bZ,\bz;\lambda_0)-\te\right)}\\
&= e^{-\epsilon \left(
H_{tot}(\bZ,\bz;\lambda_0)-\te\right)}\delta\left(H_{tot}(\bZ,\bz;\lambda_0)-\te\right)  \\
&=\delta\left(H_{tot}(\bZ,\bz;\lambda_0)-\te\right),
\end{align}
In the complex plane this contour passes parallel to the real axis,
and is shifted down by an amount $\epsilon$. 
One could reach the same result by noting that the
integrand in \eqref{deltam} is an analytical function everywhere and thus the integration contour can be
shifted down without changing the value of the integral.
We will denote this contour by $\cone$ and use
\begin{align} 
\label{eq:delta}
\delta(H_{tot}(\bZ,\bz;\lambda_0)-\te) = \frac{1}{2\pi} \int_{\cone}
dz\, e^{-i z (H_{tot}(\bZ,\bz;\lambda_0)-\te)}.
\end{align}

\subsubsection{Denominator of Eq.~\eqref{eq:JE}}

The denominator of Eq.~\eqref{eq:JE} gives the density of states associated with the initial Hamiltonian.
Using Eq.~\eqref{eq:delta}, we write this density as:

\begin{align}
\nonumber
  \dos_i(E) =& 
  \frac{1}{2\pi}\int_\cone dz\, e^{i z \te} \int d \bZ\, e^{-i z
  H_{sys}(\bZ;\lambda_0)}\\
  &\times \int d \bz\, e^{-iz(H_{int}(\bZ,\bz)+H_{env}(\bz))}
\end{align}

We begin by evaluating the last factor appearing above:
\begin{align}
  \nonumber
  \int d &\bz\, e^{-iz(H_{int}(\bZ,\bz)+H_{env}(\bz))}\\
  &=
  \left( \frac{2\pi}{i\omega z} \right)^N
  \exp\left( i z X^2 \sum_n\frac{c_n^2}{2 m_n
  \omega_n^2} \right) \\
  &= \left( \frac{2\pi}{i\omega z} \right)^N e^{i z M \gamma(0) X^2}
  \label{eq:intout}
\end{align}
where $\omega^N \equiv \omega_1\cdots \omega_N$.
The integrals are convergent due to the negative imaginary part
of $z$ as the contour $\cone$ is shifted below the real axis.

Using the definition of the renormalized frequency \eqref{def:omegabar} we get:
\begin{align}
\dos_i(E) &=
\frac{1}{2\pi} \left( \frac{2\pi}{i\omega} \right)^N
 \intc \frac{e^{i z \te}}{z^N}\int d \bZ\, e^{-i z H^*(\bZ;\lambda_0)} \\
 &=  \frac{1}{i\bar\Omega}  \left( \frac{2\pi}{i\omega} \right)^N 
 \intc \frac{e^{i z \aa}}{z^{N+1}} \quad ,
  \label{eq:int1} 
\end{align}
where in the last equality we used the definition of $\aa$ introduced in Eq.~\eqref{def:aa2}.
The sign of $\aa$ will play an important role later in the derivation. 
\begin{align} 
  \aa  =& E + \fj (\lambda) - \frac{f(\lambda)^2}{2 M \obs} =
  H_{tot}(\bZ,\bz;\lambda) + \fj (\lambda) - \frac{f(\lambda)^2}{2 M \obs} \\
   \nonumber
   =& \frac{P^2}{2 M} +\frac{1}{2} M \obs
  \left( X-\frac{f(\lambda)}{M \obs} \right)^2 \\
  &+\sum_{n=1}^{\nb}
  \left[ \frac{p_n^2}{ 2 m_n} + \frac{1}{2} m_n \omega_n^2 \left(
  x_n-\frac{c_n}{m_n \omega_n^2}X
  \right)^2 \right] \geq 0.
  \label{}
\end{align}
$\aa=0$ occurs only for a single point in the phase space. 
In the rest of this paper we take $\aa>0$.
The integral in Eq.~\eqref{eq:int1} can be evaluated by enclosing the residue at the origin,
\begin{align}
\intc \frac{e^{i z \aa}}{z^{N+1}} =  \frac{2 \pi i}{N!} i^N \aa^N. 
\end{align}
which finally brings us to the expression:
\begin{align} 
\dos_i(E) = \frac{1}{N!} \frac{(2\pi)^{N+1}}{\bar\Omega \, \omega^N} \, \aa^N .
  \label{eq:denum}
\end{align}

\subsubsection{Numerator of Eq.~\eqref{eq:JE}}
We begin by using Eq.~\eqref{eq:delta} to express the numerator as follows:
\begin{widetext}
\begin{align} 
  \nonumber
  e^{-\bbar \Delta \fj} \frac{1}{2\pi}&\intc e^{i z \te} \int d \bZ\, e^{-i z
  H_{sys}(\bZ;\lambda_0)} \int d \bz\,
  e^{-i z (H_{int}(\bZ,\bz)+H_{env}(\bz))}\\
  & \hspace{15mm} \times e^{\bbar\int_0^\tau dt\, \dot{f}(\lambda_t)
  \left[ X h(t) + P g(t) +\int_0^t ds\, g(t-s)(f(\lambda_s)-2M X \gamma(s)+\xi(s)) \right]}
  \label{}\\
  \nonumber
  =& e^{-\bbar \Delta\fj}e^{\bbar\int_0^\tau dt\, \int_0^t ds\,
  \dot{f}(\lambda_t)g(t-s)f(\lambda_s) }\frac{1}{2\pi} \intc
  e^{i z \te} \\
  \nonumber
  &\hspace{15mm} \times \int d\bZ\, e^{-i z H_{sys}(\bZ;\lambda_0)+X \left[ \bbar
  \int_0^\tau dt\,
  \dot{f}(\lambda_t) h(t)- 2 M \bbar \int_0^\tau dt\, \int_0^t ds\,
  \dot{f}(\lambda_t)
  g(t-s) \gamma(s) \right] +P \bbar \int_0^\tau dt\, \dot{f}(\lambda_t)g(t)}\\
  &\hspace{15mm}\times \int d \bz\,
  e^{-iz(H_{int}(\bZ,\bz)+H_{env}(\bz))+\bbar\int_0^\tau dt\, \int_0^t ds\,
  \dot{f}(\lambda_t) g(t-s) \sum_n c_n \left( x_n \cos(\omega_n
  s)+\frac{p_n}{m_n \omega_n}\sin(\omega_n s) \right)}.
  \label{eq:long}
\end{align}
\end{widetext}
To simplify the notation we define
\begin{align}  
  \phi_n &= c_n \bbar \int_0^\tau dt\, \int_0^t ds\,  \dot{f}(\lambda_t) g(t-s)
  \cos(\omega_n s),\\
\psi_n &= c_n \bbar \int_0^\tau dt\, \int_0^t ds\, \dot{f}(\lambda_t) g(t-s)
  \sin(\omega_n s),\\
  I_h &= \int_0^\tau dt\, \dot{f}(\lambda_t) h(t), \\
  I_g &= M\bar{\Omega}\int_0^\tau dt\, \dot{f}(\lambda_t) g(t).
 \end{align}
After integration by parts the second factor of the first line of Eq.~\eqref{eq:long} can be rewritten as
\begin{align} 
  e^{\bbar \int_0^\tau dt\, \int_0^t ds\,
  \dot{f}(\lambda_t)g(t-s)f(\lambda_s)
  }=e^{\bbar \fg - \frac{\bbar}{M \obs}I_f - \frac{\bbar}{M
  \obs}f(\lambda_0)I_h},
\end{align}
where $I_f$ has been defined in \eqref{eq:I_f2}. 
The last Gaussian integral over $\bz$ in Eq.~\eqref{eq:long}
yields:
\begin{align}  
  \left(\frac{2 \pi}{i \omega z}\right)^N e^{ i z X^2 \sum_n\frac{c_n^2}{2 m_n \omega_n^2} + X
  \sum_n\frac{c_n \phi_n}{m_n
  \omega_n^2}-\frac{i}{z} \sum_n \frac{\phi_n^2+\psi_n^2}{2 m_n
  \omega_n^2}}.
  \label{eq:3exp}
\end{align}
The first term in the exponent above can be added to
$-i z H_{sys}(\bZ;\lambda_0)$ on the second line of
Eq.~\eqref{eq:long} to give $-iz
H^*(\bZ;\lambda_0)$. The second
term in the exponent of Eq.~\eqref{eq:3exp} can be shown to be equal to:
\begin{align} 
 X \sum_n\frac{c_n \phi_n}{m_n
  \omega_n^2} = X \bbar 2 M \int_0^\tau dt\, \int_0^t ds\, \dot{f}(\lambda_s) g(t-s)
  \gamma(s).
  \label{}
\end{align}
This term cancels the corresponding term on the second line of
Eq.~\eqref{eq:long}. 

The third term of the sum in the exponent of 
Eq.~\eqref{eq:3exp} is independent of $\bZ$ and can be pulled out of
the $\bZ$ integration. Using the definitions of $\phi_n$ and $\psi_n$ it
can also be written as
\begin{align}
  \nonumber
 \bb\equiv& \sum_n \frac{\phi_n^2+\psi_n^2}{2 m_n \omega_n^2} \\ 
 \nonumber
 =&  M \bbar^2
  \int_0^\tau dt\, \int_0^\tau dt'\, \dot{f}(t)
  \dot{f}(t')\int_0^t ds\,  \\
  &\times \int_0^{t'} ds'\,  g(t-s)\gamma(s-s')
  g(t'-s').
  \label{eq:quadint}
\end{align}
In Appendix~\ref{app:doublelaplace} it is shown that the expression
for $\bb$ can be simplified further by using the relations \eqref{eq:HGRel} to obtain:
\begin {align}
\label{eq:quadint2}
\nonumber
  2 \int_0^t ds\, &\int_0^{t'} ds'\,  g(t-s)\gamma(s-s') g(t'-s')\\ 
  &=
  \frac{h(|t-t'|)}{M^2 \obs}-\frac{h(t) h(t')}{M^2 \obs}-g(t)g(t'),\\
  \bb &= \frac{\bbar^2}{2 M \obs} \left(2 I_f-I_h^2-I_g^2 \right).
  \label{eq:B}
\end{align}
The factor of two in front of $I_f$ is due to the fact that both
integration limits are from $0$ to $\tau$ in Eq.~\eqref{eq:quadint}
whereas the second integral is from $0$ to $t$ in Eq.~\eqref{eq:I_f2}.

Note that $\bb\ge 0$, which can be seen from its definition \eqref{eq:quadint}. Together with \eqref{eq:B} this indicates that
$I_f\ge 0$. This fact will soon be used in the following derivation. 

The $\bZ$ integration of Eq.~\eqref{eq:long} yields:
\begin{align}
  \nonumber
  \int d\bZ\, &e^{-i z H^*(\bZ,\lambda_0)+X \bbar I_h} \\ 
  &=\frac{2
  \pi}{i \bar{\Omega} z} e^{-i \frac{(\bbar I_h+i z f(\lambda_0))^2}{2 z M
  \obs}-iz \fj(\lambda_0)-\frac{i}{z}\frac{\bbar^2 I_g^2}{2 M \obs}}.
  \label{}
\end{align}
Gathering all the terms Eq.~\eqref{eq:long} becomes, after a number of
cancellations: 
\begin{align}  
  \frac{(2\pi)^{N}}{i^{N+1}\omega^N \bar{\Omega}} e^{-\bbar \Delta\fj} e^{\bbar \fg - \frac{\bbar}{M\obs} I_f} \intc
 \frac{ e^{i z \aa
  -\frac{i}{z} \frac{\bbar^2}{ M \obs}I_f} }{z^{N+1}}. 
  \label{eq:above}
\end{align}

In order to proceed further we have to treat two cases separately:
$I_f>0$ and $I_f=0$.
For the more general case $I_f>0$ we define $\dd$ as in
\eqref{def:dd2}
and change the integration
  variable to  $z\rightarrow z\sqrt{\dd/\aa}$, where $\dd/\aa>0$ as
  explained before. Then the integral
  becomes
  \begin{align}  
    \left( \frac{ \aa}{\dd} \right)^{N/2}\intc \frac{e^{i
    \sqrt{\aa \dd}\left( z-\frac{1}{z}
    \right)}}{z^{N+1}}.
    \label{eq:bessel}
  \end{align}
  which is proportional to a Bessel function of second
  kind: $(2 \pi i) J_N(i 2\sqrt{\aa \dd})= (2 \pi i) i^N I_N(\sqrt{4
  \aa \dd})$. Therefore, 
  the numerator in Eq.~\eqref{eq:JE} can finally be written as 
  \begin{align}   
    e^{-\bbar \Delta \fj} e^{\bbar \fg - \bbar^{-1}\dd}
    \frac{(2\pi)^{N+1}}{\omega^N \bar{\Omega}}  \frac{\aa^{N/2}}{\dd^{N/2}} I_N(\sqrt{4\aa \dd}).
    \label{eq:num}
  \end{align}
Combining \eqref{eq:denum} and \eqref{eq:num} we obtain the final
result \eqref{eq:jefn}.

In the $I_f=0$ case the integral in \eqref{eq:above} is identical to Eq.~\eqref{eq:int1}, and cancels that term in Eq.~\eqref{eq:JE} to yield:
\begin{align} 
\langle e^{-\bbar W} \rangle_{mc} = e^{-\bbar \Delta F}.
\label{eq:I_f=0JE}
\end{align}
Note that this is true for any choice of $\bbar$ irrespective of the
total energy $\te$. As is shown in Sec.~\ref{sec:P(W)}, this is a
consequence of the fact that $I_f=0$ corresponds to a delta function
work distribution at $W=\Delta F$. 
One example of this case is given in Ref.~\cite{Jarzynski2007}. For
realistic environments we expect $I_f>0$.\footnote{$I_f\ge 0$ is a
direct consequence of and can be proven directly using the fact that
$\gamma(t)$ is a positive function. 
Realistic environments will most likely be described by strictly positive dissipation kernels which in turn yield the strict inequality $I_f>0$ via application of Bohner's theorem~\cite{Bhatia2006}. The reason behind this is that for an environment with strictly positive dissipation kernel the average dissipated energy is always positive, whereas for a positive dissipation kernel it is possible that after a while all the dissipated energy, but not more, can flow back into the system. 
For any finite $\nb$ and arbitrarily large $\tau$ this is certainly the case. But for large environments and realistic $\tau$ we expect this special case to be very improbable.
$I_f<0$ can not occur in our model as mentioned before, which is due to
the fact that the harmonic oscillator environment is a passive
environment.}

Finally we note that the result of $I_f=0$ case, i.e. \eqref{eq:I_f=0JE}, can be recovered from that of $I_f>0$ case by taking the limit $I_f\rightarrow 0$ (or equivalently $\dd \rightarrow 0$) in \eqref{eq:jefn} and using the asymptotic formula $I_N(x)\approx x^N/ N! 2^N$ as $x\rightarrow 0$.
Thus Eq.~\eqref{eq:jefn} is valid for the most general case $I_f \ge 0$.

\vspace{-3mm}
\section{Derivation of Eq.~\eqref{eq:quadint2}} \label{app:doublelaplace}

To derive Eq.~\eqref{eq:quadint} we follow the method described in~\cite{Mai2007}. First observe that like any even function the damping
kernel can be written as: $\gamma(t) = \gamma(t)
\theta(t) + \gamma(-t) \theta(-t)$. We substitute this form into
Eq.~\eqref{eq:quadint} and then take Laplace transforms with respect to
$t$ and $t'$ denoted by the operators $\mathcal{L}_t(z)$ and
$\mathcal{L}_{t'}(z')$ respectively.
\begin{align}
  \nonumber
  \mathcal{L}_{t'}&(z')\mathcal{L}_{t}(z) \\
  \nonumber 
  &\times \bigg\{ 2\int_0^t\hspace{-0.3mm} ds\,
  \int_0^{t'}\hspace{-0.3mm} ds'\,  g(t\hspace{-0.3mm}-\hspace{-0.3mm}s) \gamma(s\hspace{-0.3mm}-\hspace{-0.3mm}s')\theta(s\hspace{-0.3mm}-\hspace{-0.3mm}s') g(t'\hspace{-0.3mm}-\hspace{-0.3mm}s')\\
  &+ 2\int_0^t\hspace{-0.3mm} ds\,
  \int_0^{t'}\hspace{-0.3mm} ds'\, g(t\hspace{-0.3mm}-\hspace{-0.3mm}s) \gamma(s'\hspace{-0.3mm}-\hspace{-0.3mm}s)\theta(s'\hspace{-0.3mm}-\hspace{-0.3mm}s) g(t'\hspace{-0.3mm}-\hspace{-0.3mm}s')  \bigg\}.
  \label{eq:app1}
\end{align}
Let us consider the first term. If we treat
$\gamma(s-s')\theta(s-s')$ as a function of $s$ only, the Laplace
transform with respect to $t$ has the form of a convolution of this
function with $g(t-s)$. The result is the product of Laplace
transforms of each function. Using the formula for the Laplace transform of
time-shifted functions:
\begin{align}  
  \mathcal{L}_t(z) \left\{ f(t-a)\theta(t-a) \right\} = e^{-a z}
  \hat{f}(z),
  \label{}
\end{align}
we get for the first term of Eq.~\eqref{eq:app1}:
\begin{align}  
  \mathcal{L}_{t'}(z')\hspace{-0.5mm}\left\{\hspace{-0.5mm} 2\int_0^{t'}\hspace{-1.9mm}
  ds'\,\hspace{-0.5mm} g(t'\hspace{-0.5mm}-\hspace{-0.5mm}s') e^{-z s'}\hspace{-0.3mm}
  \hat{g}(z) \hat{\gamma}(z)\hspace{-0.5mm} \right\}\hspace{-0.5mm} =\hspace{-0.5mm} 2\frac{\hat{g}(z)\hat{g}(z')}{z+z'}
  \hat{\gamma}(z).
  \label{}
\end{align}
An identical calculation, except for the change of the order of Laplace
transforms, gives $ 2\frac{\hat{g}(z)\hat{g}(z')}{z+z'}
  \hat{\gamma}(z')
  $ for the second term of Eq.~\eqref{eq:app1}. To write the final
  answer independent of the damping kernel we use Eq.~\eqref{eq:HGRel0}
  to express $\hat{\gamma}$ in terms $\hat{g}$ and $\hat{h}$.
  \begin{align} 
    \nonumber
    \frac{\hat{g}(z)\hat{g}(z')}{z+z'}&\left( \frac{\hat{h}(z)}{M
    \hat{g}(z)}+\frac{\hat{h}(z')}{M
    \hat{g}(z')}-(z+z') \right) \\
    =&
    \frac{\hat{h}(z)\hat{g}(z')+\hat{g}(z) \hat{h}(z')}{M
    (z+z')}-\hat{g}(z)\hat{g}(z').
    \label{}
  \end{align}
  Then write the first term exclusively in terms of $\hat{h}$ again
  using Eq.~\eqref{eq:HGRel0}, i.e. $\hat{g}(z)=\left( 1-z
  \hat{h}(z) \right)/ M \obs$.
  \begin{align} 
    \frac{1}{M^{2}\obs} \frac{\hat{h}(z) + \hat{h}(z')}{
    (z+z')}-\frac{\hat{h}(z)\hat{h}(z')}{M^{2}\obs}-\hat{g}(z)\hat{g}(z').
    \label{eq:befinv}
  \end{align}
  Using $\mathcal{L}_t(z)\mathcal{L}_{t'}(z')\left\{
  \frac{\hat{h}(z)}{z+z'} \right\} = \mathcal{L}_{t}(z) \left\{
  e^{-t' z}\hat{h}(z)
  \right\} = h(t-t') \theta(t-t')$, it is easily verified that the double inverse Laplace transform of
  Eq.~\eqref{eq:befinv} proves Eq.~\eqref{eq:quadint}.

\acknowledgments

The authors would like to thank Dr. Yury A. Brychkov for the proof of
Eq.~\eqref{eq:diflim} as outlined in the footnote.
C.J. acknowledges support from the National Science Foundation (USA) under grant DMR-1206971.
Y.S. is grateful to Perimeter Institute for
Theoretical Physics for their hospitality, where part of this work has been done.

\bibliography{microcanonical}{}
\bibliographystyle{apsrev4-1}

\end{document}